\documentclass{spie}
\usepackage[latin9]{inputenc}
\usepackage{amsfonts}
\usepackage{color}
\usepackage{bm}
\usepackage{amsmath}
\usepackage{amssymb}
\usepackage{graphicx}
\usepackage[unicode=true,
 bookmarks=false,
 breaklinks=false,pdfborder={0 0 1},backref=section,colorlinks=true]
 {hyperref}
\hypersetup{
 allcolors=blue}
\usepackage{authblk}

\makeatletter

\providecommand{\tabularnewline}{\\}

\everymath{\displaystyle}

\title{Innovative schemes for Correlation Plenoptic Imaging}

\author[a]{Gianlorenzo Massaro}
\author[b]{Francesco Di Lena}
\author[a,b]{Augusto Garuccio}
\author[a,b]{\\ Francesco V. Pepe}
\author[a,b]{Milena D'Angelo}

\affil[a]{Dipartimento Interateneo di Fisica, Università degli studi di Bari, I-70126 Bari, Italy}
\affil[b]{INFN, Sezione di Bari, I-70125 Bari, Italy}

\authorinfo{Send correspondence to Francesco V. Pepe. E-mail: francesco.pepe@ba.infn.it}

\makeatother

\begin{document}
\maketitle
\begin{abstract}
CPI is a novel imaging modality capable of addressing the intrinsic
limitations of conventional plenoptic imaging - namely, the resolution
loss and the sacrificed change of perspective, - while guaranteeing
the typical advantages of plenotpic imaging: 3D imaging, refocusing
of acquired pictures, in post-processing, and depth of field extension.
In this work, we review a recently developed CPI scheme, named \textit{correlation
plenoptic imaging between arbitrary planes} and derive the algorithm
for image refocusing.
\end{abstract}

\keywords{3D imaging, correlation imaging, quantum imaging, plenoptic imaging}

\section{INTRODUCTION}
\label{sec:intro} 

Plenoptic imaging (PI) allows measuring the three-dimensional light-field
distribution within a single exposure. \textit{Plenoptic} information
simultaneously encodes both the \textit{spatial} intensity distribution
impinging on a sensor, which the only data available when performing
standard imaging, and \textit{directional} information about how light
propagates in the scene\cite{PI,PI1}. The combined availability of
spatial and directional information makes it possible to reconstruct
the light paths in post-processing, unfolding a set of interesting
possibilities such as depth-of-field (DOF) extension, refocusing,
altering the point of view. PI is also one of the most convenient
methods to-date to perform scanning-free single-shot 3D reconstruction\cite{Broxton:13,Xiao:13,PI6,c72b58c6767e455488b7dd420d55b06a}. 

Conventionally, plenoptic imaging collects both spatial and directional
information on a single sensor by employing a micro-lens array\cite{PI2,PI3,PI4},
placed between the detector and an otherwise standard optical apparatus;
the presence of the micro-lenses, although necessary for multi-perspective
imaging and DOF extension, drastically reduces the minimum resolution
attainable\cite{5989827}. This translates into a marked trade-off
between DOF and resolution. Attempts have been made to improve such
resolution vs DOF issue\cite{PI5,PI6,PI7}, though it is the presence
of the array itself that imposes an intrinsic limitation to the best
resolution attainable. With the aim of completely overcoming the limitation,
thus enabling diffraction-limited imaging in a plenoptic apparatus,
we proposed \textit{correlation plenoptic imaging} (CPI)\cite{firstCPI}.

The main idea behind CPI is to decouple the two types of information
that determine the resolution and DOF properties of the final image,
performing simultaneous measurements on two separate sensors. In fact,
the DOF vs resolution trade-off arises from collecting both directional
information, which determines the DOF, and spatial information, which
determines the resolution, on the same sensor. Separating the two
measurements onto two distinct detectors allows not to sacrifice any
longer one in favor of the other\cite{PhysRevLett.119.243602}. CPI
works equally well by exploiting the spatio-temporal correlation properties
of entangled photon pairs, generated by Spontaneous Parametric Down-Conversion
(SPDC)\cite{CPIent,overview} or the statistical properties of chaotic
light\cite{overview,article}.

In Section \ref{sec:CPI-AP} a recently proposed CPI scheme named
\textit{correlation plenoptic imaging between arbitrary planes} (CPI-AP)\cite{CPI-AP}
will be presented while Section \ref{sec:refocus} will be devoted
to explaining how the focused image of a 2D object can be recovered
even when the object is not placed on the conjugate plane of either
of the detectors.

\section{CORRELATION PLENOPTIC IMAGING \\ BETWEEN ARBITRARY PLANES}
\label{sec:CPI-AP}

\begin{figure}
\begin{centering}
\begin{tabular}{c}
\includegraphics[width=0.8\textwidth]{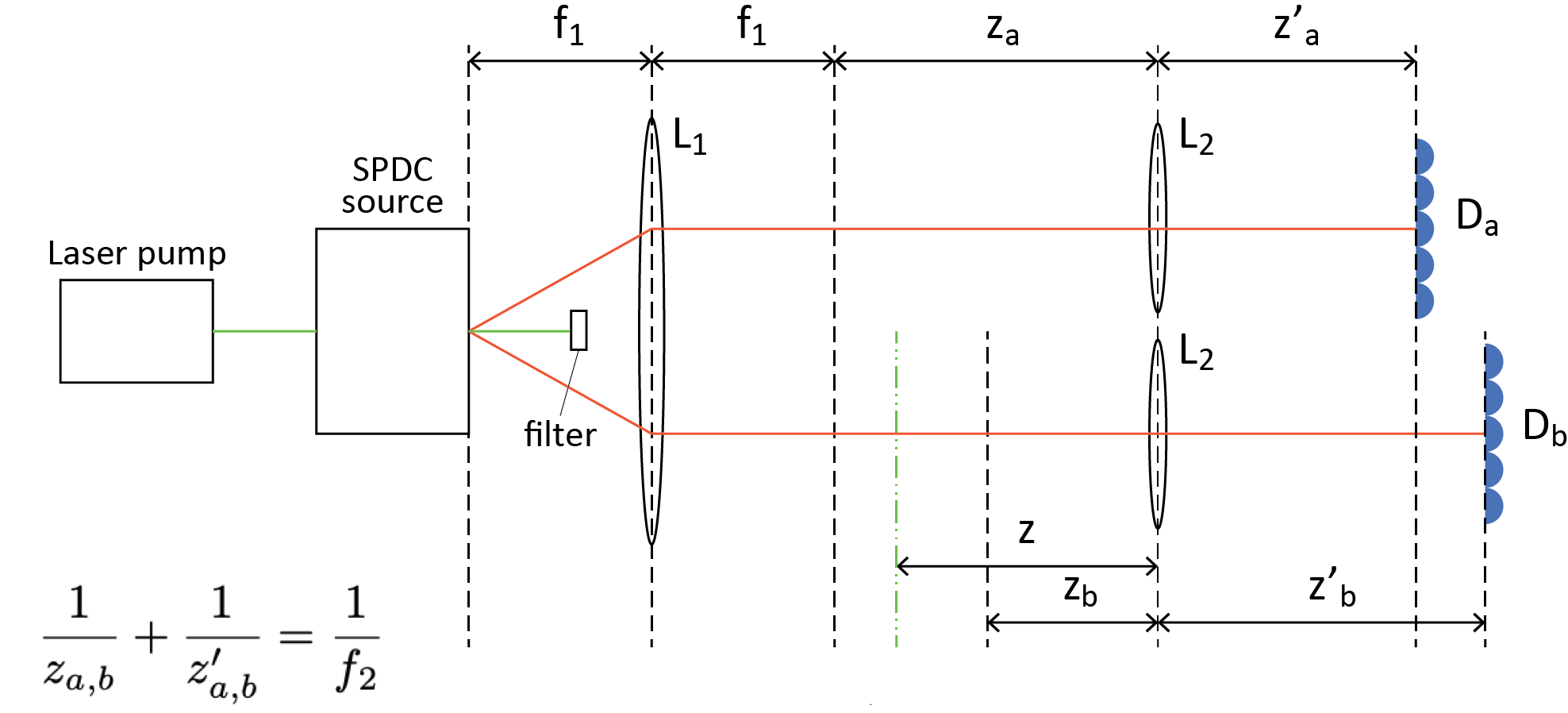}\tabularnewline
\end{tabular}
\par\end{centering}
\caption[Setup]{ \label{fig:setup} Schematics for the CPI-AP protocol. Strongly
momentum-correlated photon pairs propagate simultaneously in the upper
and lower arms. Lenses $L_{2}$ image two different planes onto detectors
$D_{A}$ and $D_{B}$. The object, in green, lies on a plane placed
anywhere between $L_{1}$ and $L_{2}$ in the same arm as $D_{B}$.
Correlation measurements between the two sensors enable the retrieval
of the object aperture even when neither its focused first-order image
is available on $D_{B}$ nor its ghost image\cite{ghost} is on $D_{A}$.}
\end{figure}

Until very recently, all proposed CPI configurations had one arm allotted
to retrieving the image of a focusing element, be it a lens\cite{overview,CPIent,CPM,scala}
or the source\cite{firstCPI,overview}, in order to collect directional
information\cite{firstCPI}. A slightly different mindset suggests,
however, that such directional propagation data, that in previous
setups was obtained by tracing rays from the object plane to the lens
or source that was imaged, can still be collected when the focusing
element is not directly imaged onto a sensor. Directional information,
in fact, is still encoded inside the correlation function when it
refers to a pair of arbitrarily chosen planes.

A schematic representation of the CPI-AP setup with entangled-photons
illumination is shown in Figure \ref{fig:setup}. A laser beam illuminates
a non-linear crystal, oriented in such a way that degenerate type-II
SPDC phase-matching conditions are satisfied\cite{pittman,rubin,rubin1};
the two entangled photons that that are generated within a single
down-conversion process are strongly frequency- and momentum-correlated
and are polarized orthogonal to each other. The laser pump after
the crystal is suppressed by a filter. The two correlated beams produced
by SPDC are separated by means of a polarizing beam splitter and propagate
in each of two arms of which the setup is composed. In the upper arm,
a lens $L_{1}$ is placed a focal length away from the source surface,
so that its Fourier transform is available on the second focal plane.
Lens $L_{2}$ of focal $f_{2}$, then, images such focal plane on
the spatially resolving detector $\text{\ensuremath{D_{a}}}$.\\
 In the lower arm, the same lens $L_{1}$ Fourier-transforms the source
light distribution in its second focal plane. A second lens $L_{2}$,
identical to the one in the other arm is still placed at distance
$z_{a}$ from the second focal plane, but the sensor $D_{b}$ is shifted
so that it collects the focused image of a different plane than the
second focal plane of $L_{1}$. The object (in green) can be placed
anywhere between $L_{1}$ and $L_{2}$, at a distance $z$ from the
latter.

CPI-AP further enhances the maximum achievable DOF of CPI and, if
the detectors are placed in such a way that $z_{A}-z_{B}\sim\text{DOF}$,
the area around the two planes conjugate to the detectors has been
proved to possess very interesting resolution and DOF properties;\cite{CPI-AP} the ability of imaging a stack
of planes along the optical axis at near-diffraction-limited resolution
within a single measurement makes CPI-AP very promising for use in
microscopy. The unprecedented DOF vs resolution property is probably
the most impressive of the advantages that the transition from the
previous CPI protocols\cite{firstCPI,overview,scala,CPM} to CPI-AP
offers; for example, precise focusing of the source or a lens, a requirement
that has proved to be rather critical to accomplish experimentally,
is not involved. It is expected that this will not only simplify certain
practical issues encountered when aligning and calibrating the apparatus,
but it will also remove ambiguities due to the non-ideal optical components in formulating the algorithms for image-recovery.

The plenoptic information in CPI is contained in the equal-time second-order
correlation function
\begin{equation}
G^{(2)}(\bm{\rho}_{a},\bm{\rho}_{b})=\left\langle \psi\right|E_{a}^{(-)}(\bm{\rho}_{a})E_{b}^{(-)}(\bm{\rho}_{b})E_{b}^{(+)}(\bm{\rho}_{b})E_{a}^{(+)}(\bm{\rho}_{a})\left|\psi\right\rangle ,\label{eq:corrfunc}
\end{equation}
where $\bm{\rho}_{a}$ and $\bm{\rho}_{b}$ are the transverse coordinates
on detectors $D_{a}$ and $D_{b}$ respectively and $E_{i}^{(\pm)}$
are the negative- and positive-frequency terms of the quantum electric
field operator; the subscript $i=a,b$ is a reminder that the field
operators in the two arms are orthogonally polarized. The expectation
value is taken over the so called ``biphoton'' state

\begin{equation}
\left|\psi\right\rangle =c_{0}\left|0\right\rangle +c_{2}\int d^{2}\kappa_{s}d^{2}\kappa_{i}\ h_{tr}(\bm{\kappa}_{s}+\bm{\kappa}_{i})\hat{a}_{k_{s}}^{\dagger}\hat{a}_{k_{i}}^{\dagger}\left|0\right>,\label{eq:biph}
\end{equation}
where $h_{\text{tr}}$ stands for the Fourier transform of laser pump.
The first term of Eq. (\ref{eq:biph}) does not give any contribution
to the correlation function, while the second term is the superposition
of momentum-correlated entangled photon pairs. The role of $h_{\text{tr}}$
is crucial, in the sense that the wider the laser pump, the stronger
the mode correlation between photons in the same pair; in the limit
$h_{\text{tr}}\left(\bm{\kappa}\right)\rightarrow\delta^{(2)}\left(\bm{\kappa}\right)$,
there is perfect mode correlation between a photon propagating in
the upper arm with momentum $\bm{\kappa}$ and its ``twin'' propagating
in the lower arm with momentum $-\bm{\kappa}$. Such a strong mode
correlation is converted into a strong point-to-point correlation
between the two arms by the Fourier-transforming property of lens
$L_{1}$. If $g_{i}\left(\bm{\kappa},\bm{\rho}\right)$, $i=a,b$
are the upper and lower arm Green's functions propagating the mode
$\bm{\kappa}$ emitted from the source to the point $\bm{\rho}$ of
the detector, Eq.~(\ref{eq:corrfunc}) reads
\begin{equation}
G^{(2)}(\bm{\rho}_{a},\bm{\rho}_{b})=\left|\int d^{2}\kappa_{a}d^{2}\kappa_{b}\ g_{a}(\bm{\kappa}_{a},\bm{\rho}_{a})g_{b}(\bm{\kappa}_{b},\bm{\rho}_{b})h_{tr}(\bm{\kappa}_{a}+\bm{\kappa}_{b})\right|^{2}.\label{eq:corrCPI}
\end{equation}
All the information about the object aperture function, which will
be indicated by $A\left(\bm{\rho}\right)$, is encoded in $g_{b}\left(\bm{\kappa},\bm{\rho}\right)$,
which also describes all the actions on the electric field determined
by optical components. The lens positions, for example, affect the
point-to-point correspondence between planes in the two arms, while
it can be shown that the finite apertures, although playing an important
role in determining the resolution, do not have an influence on determining
such point-to-point correspondence. Recalling that the pump finite
width causes a less than perfect mode-to-mode correspondence, both
the apertures and the function $h_{tr}$ play a very similar role
in smearing punctual correlations, ultimately determining the point
spread function of the system.\\
The point-to-point correspondence between the detectors can be obtained
by choosing infinite apertures and laser pump in Eq.~(\ref{eq:corrCPI}),
so that the finite size of the optical components do not influence
the correlation function, and then by performing a stationary phase
approximation (geometrical optics limit) in the remaining integrals,
yielding
\begin{equation}
G^{(2)}(\bm{\rho}_{a},\bm{\rho}_{b})\sim\left|A\left(\alpha\frac{\bm{\rho}_{a}}{M_{a}}+(\alpha-1)\frac{\bm{\rho}_{b}}{M_{b}}\right)\right|^{2},\label{eq:geom}
\end{equation}
with 
\begin{equation}
\alpha=\frac{z-z_{b}}{z_{a}-z_{b}}, \qquad M_{i}=\frac{z'_{i}}{z_{i}}.
\end{equation}
This expression shows that, when the effects of diffraction can be
neglected, the 2D correlation function at fixed coordinate on one
of the two detectors is a shifted and rescaled image of the object
transmissivity. Changing the fixed coordinate results in a shift in
the pattern on the other detector. This clearly explains the multi-perspective
imaging and change of the point of view capabilities of CPI.

When 3D imaging is not required, the SNR of the acquired images can
be significantly improved by stacking together multiple
images of the object corresponding to different points of view in
the correlation function. From Eq. (\ref{eq:geom}), it is easily
conveyed how integrating over one of the two detector planes would
result in a uniform picture\footnote{This is true unless $G^{(2)}(\bm{\rho}_{a},\bm{\rho}_{b})$ loses
its dependence on one of the two variables. This happens when $\alpha=0,1$,
namely when the object is placed on the plane available on $D_{A}$
through ghost imaging or on the plane conjugate at first order to
$D_{B}$. In this two cases, integrating over the other detector yields
the desired signal-to-noise ratio enhancement}, hence in the complete loss of information about the object. Transforming
Eq. (\ref{eq:geom}) in a way that makes stacking possible without
loss of information is what is referred to as refocusing.\cite{PI,PI1,PI4}

At this point, it is worth observing that, although we only deal with
entangled-photon illumination here, the principle of CPI-AP is easily
extended to the case of chaotic light,\cite{CPI-AP}
by measuring the correlations of intensity fluctuations $\left\langle I(\bm{\rho}_{a})I(\bm{\rho}_{b})\right\rangle -\left\langle I(\bm{\rho}_{a})\right\rangle \left\langle I(\bm{\rho}_{b})\right\rangle $,
with the angle brackets denoting the ensemble average on the quantity
enclosed. In this case, since two copies of the speckle pattern\cite{scala}
generated by the source are obtained through a beam splitter, the
mode correspondence between the two arms is of the kind $\bm{\kappa}\leftrightarrow\bm{\kappa}$,
in place of $\bm{\kappa}\leftrightarrow-\bm{\kappa}$ typical of the
entangled case, resulting in only a slight modification in the refocusing
algorithm that is going to be shown in the next section.

\section{REFOCUSING ALGORITHM}

\label{sec:refocus}

We define the refocusing function as
\begin{equation}
\Sigma\left(\bm{\rho}_{r}\right)=\int d^{2}\rho_{s}\ G^{(2)}\left(\bm{\rho}_{r},\bm{\rho}_{s}\right)\label{eq:refFunc}
\end{equation}
where $\bm{\rho}_{r}$ and $\bm{\rho}_{s}$ are obtained by transformation
of the original coordinates at the detector. The choice of the new
set of coordinates has to be done in such a way to obtain the focused
image of the object $\left|A\left(\bm{\rho}_{s}\right)\right|^{2}$
when integration over $\bm{\rho}_{s}$ is carried over. From Eq. (\ref{eq:refFunc}),
it is clear that a suitable parametrization for the coordinate $\bm{\rho}_{r}$
would be
\begin{equation}
\bm{\rho}_{r}=C_{11}\bm{\rho}_{a}+C_{12}\bm{\rho}_{b}\label{eq:rhor}
\end{equation}
with
\begin{equation}
C_{11}=\frac{\alpha}{M_{a}},\ \ \ C_{12}=\frac{\alpha-1}{M_{b}};\label{eq:coeff}
\end{equation}
we define $\bm{\rho}_{s}$ by introducing two arbitrary coefficients
$C_{21}$ and $C_{22}$, so that
\begin{equation}
\bm{\rho}_{s}=C_{21}\bm{\rho}_{a}+C_{22}\bm{\rho}_{b}.\label{eq:rhos}
\end{equation}
As we see, independent of Eq. (\ref{eq:rhos}) we have $G^{(2)}\left(\bm{\rho}_{r},\bm{\rho}_{s}\right)\sim\left|A\left(\bm{\rho}_{r}\right)\right|^{2}$;
the integrated variable $\bm{\rho}_{s}$, then, does not look like
playing a role in refocusing itself, but can be chosen in the most
convenient way. For convenience, the coefficients $C_{ij}$, $i,j=1,2$
in Eqs.~(\ref{eq:rhor})-(\ref{eq:rhos}) will be thought as the
elements of a matrix $C$ transforming the first set of coordinates
into the new ones.

\begin{figure}
\begin{centering}
\begin{tabular}{c}
\includegraphics[width=1\textwidth]{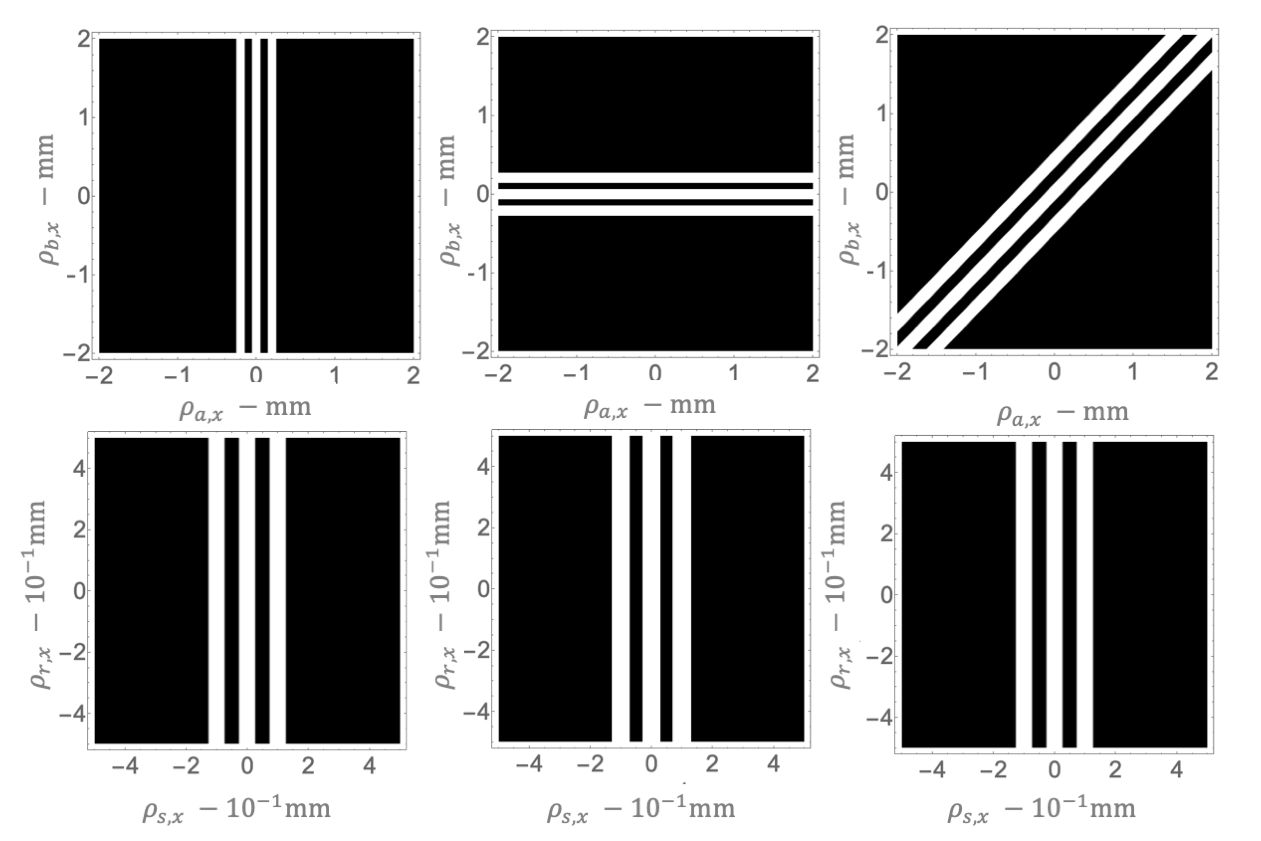}\tabularnewline
\end{tabular}
\par\end{centering}
\caption{Refocusing capabilities of the algorithm in
Eq. \eqref{eq:algorithm}. The object is a triple slit placed at $z=z_{a}$
(left), $z=z_{b}$ (center) and $z=\left(z_{a}+z_{b}\right)/2$ (right).
The lower plots are the 2D projections $G^{(2)}\left(\rho_{a,x},0,\rho_{b,x},0\right)$
of the 4D correlation function measured by the detectors. If $z=z_{a}$
or $z=z_{b}$, the focused image of the object is available on one
detector; changing ``point of view'' on the other detector does
not result in a shifting of the pattern. Integration over one detector
does not result in loss of information about the object. In the most
general case $z\protect\neq z_{a,b}$, however, the patterns are shifted. 
The lower plots show the refocused correlation function in the transformed
$\rho_{s,x}-\rho_{r,x}$ plane. The features of the objects are available
along the $\rho_{r}$-axis, while the $\rho_{s}$-axis contains identical
``copies'' of the pattern that can be superimposed without loss
of information.}\label{fig:refocus}
\end{figure}

In order to refocus the object, $G^{(2)}(\bm{\rho}_{a},\bm{\rho}_{b})$
must be reformulated in terms of the new set of coordinates; Eq. (\ref{eq:rhor})
and (\ref{eq:rhos}) need be inverted in order to find $\bm{\rho}_{a}(\bm{\rho}_{r},\bm{\rho}_{s})$
and $\bm{\rho}_{b}(\bm{\rho}_{r},\bm{\rho}_{s})$ and replacing the
result into the correlation function. Since the coefficients of $A$
depend on the position of the object along the optical axis, that
we parametrized with $\alpha$, there may be some longitudinal positions
of the object for which Eq. (\ref{eq:rhor}) and (\ref{eq:rhos})
cannot be inverted. Therefore, the choice of $C_{21}$ and $C_{22}$,
must be made in such a way that $\det C\neq0$. The condition of not
singularity of $C$ is not enough to fix the two remaining coefficients,
so we also impose that $C$ transforms the canonical basis of the
$\left(\rho_{a},\rho_{b}\right)$ plane\footnote{Since the coefficients of the matrix $C$ are the same
for both the $x$ and the $y$ components, the $2\times 2$ matrix $C$ acts in the same way on $(\rho_{a,x},\rho_{b,x})$ and $(\rho_{a,y},\rho_{b,y})$.} into an orthogonal base for the transformed plane\footnote{This choice coincides with integrating over detector $D_{a}$ when
the object is available at first order on $D_{b}$ and, conversely,
on $D_{b}$ when the ghost image is available on $D_{a}$.} $\left(\rho_{r},\rho_{s}\right)$.The aforementioned requirements
provide the transformation
\begin{equation}
C=\left(\begin{array}{cc}
\frac{\alpha}{M_{a}} & \frac{\alpha-1}{M_{b}}\\
\frac{1-\alpha}{M_{a}} & \frac{\alpha}{M_{b}}
\end{array}\right)\label{eq:transformation}
\end{equation}
and the refocusing algorithm
\begin{equation}
\begin{cases}
\bm{\rho}_{a} & =(C^{-1})_{11}\bm{\rho}_{r}+(C^{-1})_{12}\bm{\rho}_{s}\\
\bm{\rho}_{b} & =(C^{-1})_{21}\bm{\rho}_{r}+(C^{-1})_{22}\bm{\rho}_{s}
\end{cases},\label{eq:algorithm}
\end{equation}
where $(C^{-1})_{ij}$ are the elements of the inverse of matrix
$C$,
\begin{equation}
C^{-1}=\frac{1}{2\alpha^{2}-2\alpha+1}\left(\begin{matrix}\alpha M_{a} & \left(1-\alpha\right)M_{a}\\
\left(\alpha-1\right)M_{b} & \alpha M_{b}
\end{matrix}\right),\label{eq:transformation-1}
\end{equation}
always well-defined since $2\alpha^{2}-2\alpha+1\neq0,$ $\forall\alpha$.

Figure \ref{fig:refocus} shows the working principle of the refocusing
algorithm when the object to be imaged is a triple slit. In the simulation,
the 4D correlation function is reduced to a two-variable function,
a rendition that makes it clear how, when the object is not placed
on a plane conjugate to a detector, the object features shift as the
point of view is changed. It is also shown that our requirement of
$C$ never being singular yields the desired outcome for the two critical
values $\alpha=0,1$. We should remark, however, that refocusing is
based on the assumption that Eq. (\ref{eq:geom}) is a good approximation
of the correlation function. When diffraction features become noticeable,
each one of the 2D patterns obtained by fixing a detector coordinate
show the fringes typical of a diffraction pattern.\cite{overview}
Refocusing, in this case, cannot correctly retrieve a focused image
of the object since the point-to-point correspondence on which it
is based is no longer fullfilled.

\section{CONCLUSIONS}

The refocusing algorithm presented in Section \ref{sec:refocus} is
not the only suitable choice for reconstructing the object focused
image. When the finite size of the lenses is taken into account, Eq.
(\ref{eq:geom}) is modified in such a way that the correlation function
also contains the pupil functions of the lenses as multiplying factors.\cite{CPI-AP} The presence of the pupil functions,
that are zero-valued everywhere outside of a certain radius from the
optical axis, selects the part of the object that is transmitted,
therefore detemining the field of view. A different approach to the
choice of the reefocusing algorithm would be to choose the coordinate
centered on a pupil as the integrated variable $\bm{\rho}_{s}$, so
that all the images would be stacked in accordance to their field
of view. As long as the trasnformation matrices are non-singular,
all the possible approaches involving Eq. (\ref{eq:rhor}) and (\ref{eq:coeff})
are theoretically equivalent. The best choice for an algorithm is
then ultimately dictated by convenience and revolves around experimental
aspects such as the effective apertures of the optical components,
the size of the object and the sensors and the pixel density.

CPI-AP with entangled-photons illumination is currently being inplemented
in order to exploit the very high SNR that entangled photons are known
to allow in the low-photon-flux regime\cite{Samantaray,subshot}.

\section*{ACNOWLEDGMENTS}
This work was supported by QuantERA project ``Qu3D - Quantum 3D imaging at high speed and high resolution'', by Istituto Nazionale di Fisica Nucleare (INFN) project ``PICS4ME -- Plenoptic Imaging with Correlations for Microscopy and 3D Imaging Enhancement'', and by PON ARS $01\_00141$ ``CLOSE -- Close to Earth'' of Ministero dell'Istruzione, dell'Universit\`a e della ricerca (MIUR).


\end{document}